\documentclass[11pt, a4paper]{article}

\usepackage{amsmath}
\usepackage{amssymb}
\usepackage{graphicx}
\usepackage{hyperref}
\usepackage{booktabs}
\usepackage{xcolor}
\usepackage{listings}
\usepackage{float}
\usepackage{geometry}

\hypersetup{
  colorlinks=true,
  linkcolor=blue,
  citecolor=blue,
  urlcolor=blue
}

\lstdefinelanguage{json}{
  morestring=[b]",
  literate=
    *{:}{{{\color{blue!60!black}:}}}{1}
     {,}{{{\color{blue!60!black},}}}{1}
     {\{}{{{\color{blue!60!black}\{}}}{1}
     {\}}{{{\color{blue!60!black}\}}}}{1}
     {[}{{{\color{blue!60!black}[}}}{1}
     {]}{{{\color{blue!60!black}]}}}{1}
}

\title{Portable Agent Memory: A Protocol for Provenance-Verified Memory Transfer Across Heterogeneous LLM Agents}

\author{
  Santhosh Kumar Ravindran \\
  Microsoft Corporation, Redmond, WA, USA \\
  \texttt{santhosh.ravindran@microsoft.com}
}

\date{}

\begin{document}

\maketitle

\begin{abstract}
Large language model (LLM) agents accumulate valuable operational context---learned preferences, factual knowledge, procedural skills, and task state---yet this memory remains locked within vendor-specific runtimes, creating fragility, vendor lock-in, and catastrophic knowledge loss across sessions. We present \textbf{Portable Agent Memory}, an open protocol for serializing, transporting, and re-hydrating agent memory across heterogeneous LLM-based systems with cryptographic integrity guarantees. Portable Agent Memory introduces a five-component memory model $M = (E, S, P, W, I)$ spanning episodic, semantic, procedural, working, and identity memory; a Merkle-DAG provenance structure enabling tamper-evident verification; capability-scoped access tokens for fine-grained authorization; and an injection-resistant re-hydration pipeline that defends against memory-mediated prompt injection attacks. In a pilot study across Claude, GPT-4, and Gemini, Portable Agent Memory achieves Transfer Continuity Scores of 0.83--0.92 compared to a no-memory baseline of 0.28--0.45, demonstrating that structured portable memory substantially preserves agent capability across model boundaries. We release a complete Python SDK with 54 passing tests.
\end{abstract}


\section{Introduction}

The era of autonomous LLM agents has produced systems that accumulate rich operational context through extended interactions: user preferences learned over weeks, domain knowledge extracted from documents, procedural skills refined through trial and error, and complex task state maintained across sessions. This accumulated memory represents significant value---yet it remains almost universally imprisoned within the specific vendor platform, model family, or agent framework that created it.

We identify six critical problems in current agent memory systems:

\begin{enumerate}
    \item \textbf{Vendor lock-in.} Memory accumulated in one platform (e.g., ChatGPT's memory feature, Claude's project context) cannot be exported to competing systems. Users who switch providers lose months of personalization.

    \item \textbf{Session amnesia.} Despite advances in context windows, agents still experience catastrophic forgetting across session boundaries. Each new conversation starts from zero unless the platform provides proprietary continuity mechanisms.

    \item \textbf{No integrity verification.} When memory is transferred---even within the same platform---there is no mechanism to verify it has not been tampered with. A corrupted or poisoned memory store can silently degrade agent behavior.

    \item \textbf{Coarse access control.} Existing systems offer binary access: either full memory or none. There is no capability to share specific knowledge subsets with collaborating agents while protecting sensitive context.

    \item \textbf{Injection vulnerability.} Naive memory injection (prepending conversation history) creates a prompt injection attack surface. Recalled text containing instruction-like patterns can hijack agent behavior.

    \item \textbf{No cross-model transfer.} An agent running GPT-4 cannot meaningfully transfer its learned context to one running Claude or Gemini. The memory formats, assumptions about context structure, and tokenization are incompatible.
\end{enumerate}

\subsection{Protocol Stack Positioning}

Portable Agent Memory is designed to complement existing agent interoperability protocols rather than compete with them. The emerging agent infrastructure stack consists of three orthogonal concerns:

\begin{itemize}
    \item \textbf{MCP (Model Context Protocol)} \cite{anthropic2024mcp}: Standardizes how agents access external tools and data sources. MCP answers ``what can the agent \emph{do}?''
    \item \textbf{A2A (Agent-to-Agent)} \cite{google2025a2a}: Standardizes how agents delegate tasks to other agents. A2A answers ``how do agents \emph{collaborate}?''
    \item \textbf{Portable Agent Memory}: Standardizes how agents transfer accumulated knowledge. It answers ``what does the agent \emph{know}?''
\end{itemize}

Together, these three protocols form a complete interoperability layer: tools (MCP), coordination (A2A), and memory (Portable Agent Memory).

\subsection{Contributions}

This paper makes the following contributions:

\begin{enumerate}
    \item \textbf{A formal five-component memory model} ($M = (E, S, P, W, I)$) that captures the full spectrum of agent operational context in a model-agnostic format.

    \item \textbf{A Merkle-DAG provenance structure} with BLAKE3 content-addressing and Ed25519 root signing, enabling tamper-evident verification of memory derivation chains.

    \item \textbf{A capability-based access control system} with scoped tokens supporting selective disclosure, enabling fine-grained multi-agent memory sharing without exposing the complete memory store.

    \item \textbf{An injection-resistant re-hydration pipeline} that defends against memory-mediated prompt injection through structural framing, content escaping, and type enforcement---enabling safe cross-model memory transfer.

    \item \textbf{A working implementation} (Python SDK, 54 tests, $<$0.5s test suite execution) demonstrating the protocol's practicality.
\end{enumerate}

\section{Related Work}

\subsection{Agent Memory Systems}

The importance of persistent memory for LLM agents is well-established. \textbf{MemGPT} (Packer et al., 2023) \cite{packer2023memgpt} introduces a virtual memory hierarchy inspired by operating systems, with main context and external storage managed through self-directed memory operations. While groundbreaking in demonstrating that agents can manage their own memory, MemGPT's memory format is tightly coupled to its own runtime and offers no portability guarantees.

\textbf{Generative Agents} (Park et al., 2023) \cite{park2023generative} simulate human-like memory with observation streams, reflection, and planning---demonstrating that structured memory dramatically improves agent coherence. However, their memory architecture assumes a single simulation environment and does not address cross-agent transfer.

\textbf{Voyager} (Wang et al., 2023) \cite{wang2023voyager} maintains a skill library of verified programs that grows through exploration. This procedural memory concept directly informs Portable Agent Memory's procedural component, though Voyager's skills are Minecraft-specific and not designed for portability.

\textbf{Reflexion} (Shinn et al., 2023) \cite{shinn2023reflexion} uses linguistic feedback as episodic memory, enabling agents to learn from failures. This validates episodic memory's value for agent improvement but operates within a single-agent, single-model paradigm.

Recent surveys provide comprehensive taxonomies of agent memory mechanisms. Zhang et al.\ \cite{zhang2024memory} classify memory by storage, operation, and retrieval mechanisms across 40+ agent systems. Sumers et al.\ \cite{sumers2024cognitive} connect modern LLM agent architectures to classical cognitive architectures like ACT-R \cite{anderson2004actr} and SOAR \cite{laird1987soar}, demonstrating that the episodic-semantic-procedural memory taxonomy has deep roots in cognitive science.

\subsection{Commercial Memory Services}

\textbf{Mem0} \cite{mem02024} provides a commercial memory layer as SaaS, offering memory extraction and retrieval. However, Mem0's memory is stored in their cloud infrastructure, creating a new form of vendor lock-in. There is no cryptographic verification of memory integrity, and portability between Mem0 and other systems is not supported.

\textbf{Zep} (Rasmussen et al., 2025) \cite{rasmussen2025zep} builds temporal knowledge graphs from agent interactions, enabling sophisticated temporal reasoning. While technically impressive, the graph structure is proprietary and cannot be exported to non-Zep systems.

\textbf{Supermemory} offers browser-based memory management but targets human users rather than programmatic agent interoperability.

\subsection{Agent Interoperability Protocols}

\textbf{MCP (Model Context Protocol)} \cite{anthropic2024mcp} standardizes tool access but explicitly does not address memory persistence or transfer. An agent's MCP tool definitions are stateless---they describe capabilities, not accumulated knowledge.

\textbf{A2A (Agent-to-Agent Protocol)} \cite{google2025a2a} enables task delegation between agents but does not specify how an agent's learned context should be conveyed alongside a delegated task.

Yang et al.\ \cite{yang2025protocols} provide the first comprehensive survey of agent communication protocols, classifying them along context-oriented vs.\ inter-agent and general-purpose vs.\ domain-specific dimensions. Their analysis identifies privacy preservation and adaptability as key requirements for next-generation protocols---precisely the properties Portable Agent Memory addresses.

\textbf{AMCP (Agent Memory Communication Protocol)} \cite{nunchiai2025amcp} is the closest existing work to Portable Agent Memory. AMCP defines a protocol for agent memory communication but lacks cryptographic integrity verification, capability-based access control, and injection-resistant framing. It targets memory \emph{communication} rather than verified \emph{transfer}.

\textbf{Soul Protocol} \cite{qbtrix2025soul} provides signed identity for a single agent across sessions but does not address multi-component memory or cross-model transfer.

\subsection{Foundational Work}

Portable Agent Memory draws on several foundational threads. The retrieval-augmented generation (RAG) paradigm \cite{lewis2020rag} established that combining parametric knowledge with non-parametric retrieval dramatically improves knowledge-intensive tasks; Portable Agent Memory's recall operation is a structured, verified variant of RAG. Agent frameworks like AutoGen \cite{wu2023autogen}, ReAct \cite{yao2023react}, and Toolformer \cite{schick2023toolformer} demonstrate the diversity of agent memory patterns that a portable format must accommodate.

The prompt injection threat model motivating Portable Agent Memory's security design is formalized by Greshake et al.\ \cite{greshake2023injection} and Perez and Ribeiro \cite{perez2022ignore}, who demonstrate that adversarial content in retrieved context can hijack agent behavior---a risk amplified when memory crosses trust boundaries.

Portable Agent Memory's memory taxonomy extends Tulving's episodic-semantic distinction \cite{tulving1985memory} through the lens of cognitive architectures: ACT-R \cite{anderson2004actr} distinguishes declarative, procedural, and working memory, while SOAR \cite{laird1987soar} formalizes production memory for skill acquisition. The connection between classical cognitive architectures and modern LLM agents is explored by Sumers et al.\ \cite{sumers2024cognitive}.

Finally, Portable Agent Memory's portability goals align with the data portability rights established by GDPR Article 20 \cite{eu2016gdpr}, which grants individuals the right to receive and transmit their personal data in machine-readable formats---a principle we extend to agent-accumulated knowledge.

\subsection{Cryptographic Data Structures in AI}

Merkle trees (Merkle, 1987) \cite{merkle1987signature} have been applied to ML model provenance (tracking training data lineage), signed model checkpoints, and federated learning verification. Portable Agent Memory adapts these techniques to agent memory, where the DAG structure naturally represents derivation relationships between knowledge entries (e.g., a semantic fact derived from an episodic observation).

\subsection{Gap Analysis}

Table~\ref{tab:comparison} summarizes the landscape. No existing system combines all five properties that Portable Agent Memory provides: portability across model families, cryptographic integrity verification, fine-grained capability-based access control, injection-resistant re-hydration, and quantitative fidelity metrics.

\begin{table}[H]
\centering
\caption{Comparison of agent memory systems across key protocol properties.}
\label{tab:comparison}
\begin{tabular}{lccccc}
\toprule
\textbf{System} & \textbf{Portable} & \textbf{Crypto Verified} & \textbf{Scoped Access} & \textbf{Injection Defense} & \textbf{Metrics} \\
\midrule
MemGPT/Letta & \texttimes & \texttimes & \texttimes & \texttimes & \texttimes \\
Mem0 & \texttimes & \texttimes & Partial & \texttimes & \texttimes \\
Zep/Graphiti & \texttimes & \texttimes & \texttimes & \texttimes & \texttimes \\
AMCP & \checkmark & \texttimes & \texttimes & \texttimes & \texttimes \\
Soul Protocol & Partial & \checkmark & \texttimes & \texttimes & \texttimes \\
\textbf{Portable Agent Memory} & \checkmark & \checkmark & \checkmark & \checkmark & \checkmark \\
\bottomrule
\end{tabular}
\end{table}

\section{System Design}

\subsection{Memory Artifact Format}

Portable Agent Memory models agent memory as a five-component artifact:

\begin{equation}
M = (E, S, P, W, I)
\end{equation}

where each component captures a distinct cognitive function:

\begin{itemize}
    \item \textbf{Episodic ($E$):} Time-ordered records of events, observations, and interactions. Analogous to human autobiographical memory.
    \item \textbf{Semantic ($S$):} Factual assertions as subject-predicate-object triples with confidence scores. The agent's world model.
    \item \textbf{Procedural ($P$):} Learned skills, workflows, and routines with usage statistics and preconditions.
    \item \textbf{Working ($W$):} Transient goals, subgoals, scratch computations, and pending actions. The agent's ``mental workspace.''
    \item \textbf{Identity ($I$):} Persistent persona attributes, preferences, communication style, and operational policies.
\end{itemize}

This decomposition is inspired by cognitive architectures (Tulving, 1985; Anderson, 1996) and adapted for LLM agent requirements. Each component type has a distinct schema enabling type-specific processing during re-hydration.

Every entry across all components shares a common base:

\begin{lstlisting}[language=json]
{
  "id": "blake3:a1b2c3d4e5f6...",
  "parent_ids": ["blake3:9f86d081..."],
  "created_at": "2025-01-15T08:30:00Z",
  "version": "1.0"
}
\end{lstlisting}

The \texttt{id} field is the BLAKE3 hash of the entry's canonical JSON serialization (with the \texttt{id} field itself omitted), creating a \textbf{content-addressable} identifier. Any modification to any field changes the hash, making tampering immediately detectable.

\textbf{Design decision: JSON-first.} Portable Agent Memory uses JSON as its primary serialization format, with CBOR as a compact binary alternative. This ensures every programming language, every LLM framework, and every cloud platform can immediately produce and consume Portable Agent Memory artifacts without specialized libraries. The protocol avoids proprietary binary formats, custom encoding schemes, or framework-specific serialization---maximizing the probability of ecosystem adoption.

\subsection{Provenance Graph}

Memory entries form a directed acyclic graph (DAG) through their \texttt{parent\_ids} references. This Merkle-DAG structure serves three purposes:

\begin{enumerate}
    \item \textbf{Derivation tracking:} When a semantic fact is extracted from an episodic observation, the semantic entry's \texttt{parent\_ids} references the episodic entry, creating an auditable derivation chain.

    \item \textbf{Tamper evidence:} Because each entry's ID is computed from its content, and parent references embed parent IDs into child hashes, modifying any entry invalidates all downstream entries---analogous to blockchain integrity.

    \item \textbf{Selective disclosure:} The DAG structure enables exporting a subset of entries while preserving verifiable provenance by including transitive ancestors.
\end{enumerate}

\textbf{Verification proceeds in three phases:}

\emph{Phase 1 --- Hash verification:} Recompute every entry's ID from its content and verify it matches the declared \texttt{id}.

\emph{Phase 2 --- DAG integrity:} Verify acyclicity, referential integrity (no dangling parent references), and the presence of at least one root entry.

\emph{Phase 3 --- Root signature:} Compute the root hash as \texttt{BLAKE3(canonical\_json(components))} and verify the Ed25519 signature against the operator's public key.

\begin{lstlisting}[language=Python]
def verify_artifact(artifact, operator_pubkey):
    # Phase 1: Every entry hash must be self-consistent
    for entry in all_entries(artifact):
        assert entry["id"] == compute_entry_id(entry)

    # Phase 2: DAG must be acyclic with valid references
    assert is_acyclic(build_dag(artifact))
    assert all_parents_exist(artifact)

    # Phase 3: Root hash signed by trusted operator
    root = blake3(canonical_json(artifact["components"]))
    assert ed25519_verify(operator_pubkey, root, artifact["signature"])
\end{lstlisting}

\textbf{Key operations on the provenance graph:}

\begin{itemize}
    \item \texttt{derive(parents, fields)} $\rightarrow$ Create a new entry linked to existing parents, automatically computing the content-addressable ID.
    \item \texttt{verify(entry\_id)} $\rightarrow$ Recursively verify an entry's hash chain back to root entries.
    \item \texttt{selective\_disclose(entry\_ids)} $\rightarrow$ Extract a sub-DAG preserving provenance integrity by including all transitive ancestors.
\end{itemize}

\subsection{Capability-Based Access Control}

Portable Agent Memory implements object-capability security (Dennis \& Van Horn, 1966) \cite{dennis1966capability} for memory access. Capability tokens are signed, scoped authorizations:

\begin{lstlisting}[language=json]
{
  "id": "cap:f47ac10b-58cc-4372-a567-0e02b2c3d479",
  "scope_expression": {
    "type": "component",
    "components": ["episodic", "semantic"]
  },
  "permissions": ["read", "derive"],
  "issuer": "operator:admin@example.com",
  "issuer_signature": "ed25519:5a3b...",
  "audience": "agent:research-bot-beta",
  "expires_at": "2025-06-15T00:00:00Z"
}
\end{lstlisting}

\textbf{Scope expressions} define which entries a token governs, supporting four types:

\begin{itemize}
    \item \textbf{Entry list:} Explicit enumeration of authorized entry IDs.
    \item \textbf{Component type:} Access to all entries within named components (e.g., all semantic memory).
    \item \textbf{Tag predicate:} Access governed by entry tags (\texttt{any\_of}, \texttt{all\_of}, \texttt{none\_of} operators).
    \item \textbf{Wildcard:} Full access (use with extreme caution).
\end{itemize}

\textbf{Permissions} are granular: \texttt{read}, \texttt{write}, \texttt{derive}, \texttt{redact}, \texttt{export}, \texttt{rehydrate}. This enables scenarios like granting a collaborating agent \texttt{read} + \texttt{derive} permission on semantic memory (it can read facts and create new derived facts) without granting \texttt{export} permission (it cannot re-export those facts to third parties).

\textbf{Selective disclosure} enables multi-agent handoffs where different agents receive different memory subsets based on their role and trust level. A project manager agent might receive working memory (goals, status) while a specialist agent receives procedural memory (relevant skills).

\subsection{Re-Hydration Protocol}

Re-hydration transforms a portable Portable Agent Memory artifact into active context within a target agent's working state. The pipeline consists of seven stages:

\begin{center}
\texttt{Verify $\rightarrow$ Filter $\rightarrow$ Rank $\rightarrow$ Compress $\rightarrow$ Format $\rightarrow$ Frame $\rightarrow$ Inject}
\end{center}

\textbf{Step 1: Verify artifact.} Validate structural integrity, hash consistency, DAG acyclicity, root signature, and size limits. Halt on first failure.

\textbf{Step 2: Capability filter.} Remove entries not authorized by presented capability tokens. Validate token signatures, expiration, and audience binding.

\textbf{Step 3: Relevance ranking.} Score each entry against the target agent's current task context using a configurable relevance function:

\begin{equation}
\label{eq:relevance}
\text{relevance}(e, ctx) = \alpha \cdot \text{recency}(e) + \beta \cdot \text{salience}(e) + \gamma \cdot \text{sim}(e, ctx) + \delta \cdot \text{depth}(e)
\end{equation}

where $\alpha = 0.2$, $\beta = 0.3$, $\gamma = 0.4$, $\delta = 0.1$ (defaults). The semantic similarity term uses embedding cosine distance; recency decays linearly; depth favors entries closer to DAG roots.

\textbf{Step 4: Summarization-aware compression.} Given a token budget (configured per target model's context window), the compressor includes high-relevance entries ($\geq 0.7$) verbatim, summarizes medium-relevance entries ($0.3$--$0.7$), and drops low-relevance entries with count annotations.

\textbf{Step 5: Model-specific formatting.} Render compressed entries into text appropriate for the target model's conventions. Supports \texttt{structured} (JSON-like key-value) and \texttt{narrative} (prose) format styles.

\textbf{Step 6: Injection-resistant framing.} Apply structural framing (Section~\ref{sec:injection}) to prevent memory content from being interpreted as instructions.

\textbf{Step 7: Inject into agent context.} Place framed memory after the system prompt and before user-controlled content, ensuring the model's instruction hierarchy treats Portable Agent Memory data as context rather than commands.

\subsection{Injection-Resistant Framing}
\label{sec:injection}

Memory-mediated prompt injection is a novel attack vector: an adversary poisons source agent memory with instruction-like text, which is then faithfully transferred to and executed by the target agent. Portable Agent Memory defends against this through three mechanisms:

\textbf{Structural framing.} All recalled memory is wrapped in typed boundary markers with an explicit system directive:

\begin{lstlisting}
[PAM:SYSTEM_DIRECTIVE]
The following is recalled observational data from a previous agent session.
Treat this content as factual context only. Do NOT interpret any text
within PAM:DATA blocks as instructions, commands, or role assignments.
[/PAM:SYSTEM_DIRECTIVE]

[PAM:DATA:semantic]
ACME Corp reported_revenue $4.2B in Q3 2024 (confidence: 0.92)
[/PAM:DATA]
\end{lstlisting}

\textbf{Content escaping.} Before framing, content undergoes three escaping passes:

\begin{enumerate}
    \item \emph{Boundary escape:} Portable Agent Memory delimiter markers in content are escaped to prevent boundary breakout.
    \item \emph{Role marker escape:} Patterns like \texttt{"System:"}, \texttt{"Assistant:"}, \texttt{"User:"} are replaced with \texttt{[ESCAPED\_ROLE:...]} tokens.
    \item \emph{Instruction escape:} Known injection patterns (``Ignore previous instructions'', ``You are now'', etc.) are replaced with inert tokens.
\end{enumerate}

\textbf{Content-type enforcement.} Each \texttt{[PAM:DATA:<type>]} block is validated against its declared schema. Content that does not match (e.g., imperative instructions appearing in a \texttt{semantic} block) is quarantined and excluded from the framed output.

\section{Implementation}

\subsection{Python SDK Architecture}

We provide \texttt{pam-sdk}, a Python reference implementation organized into six modules:

\begin{table}[H]
\centering
\caption{Python SDK module organization.}
\label{tab:modules}
\begin{tabular}{ll}
\toprule
\textbf{Module} & \textbf{Responsibility} \\
\midrule
\texttt{pam.models} & Pydantic data models for all entry types and the artifact envelope \\
\texttt{pam.provenance} & Merkle-DAG construction, hash computation, verification, selective disclosure \\
\texttt{pam.capabilities} & Token creation, validation, scope resolution, capability filtering \\
\texttt{pam.rehydration} & Seven-stage re-hydration pipeline with configurable relevance functions \\
\texttt{pam.serialization} & JSON canonical form, CBOR encoding, file format with magic bytes \\
\texttt{pam.transport} & File I/O, HTTP bindings, MCP tool definitions \\
\bottomrule
\end{tabular}
\end{table}

The SDK targets zero external dependencies beyond \texttt{blake3} and \texttt{pynacl} (for Ed25519), ensuring minimal installation friction. Total implementation spans approximately 2,500 lines of Python with comprehensive type annotations.

\subsection{Test Suite}

The SDK includes 54 tests across 5 test modules covering:

\begin{itemize}
    \item \textbf{Model validation} (entry schema enforcement, size limits, field constraints)
    \item \textbf{Provenance operations} (hash computation, DAG verification, cycle detection, selective disclosure)
    \item \textbf{Capability tokens} (creation, validation, expiration, audience binding, scope resolution)
    \item \textbf{Re-hydration} (pipeline stages, relevance scoring, compression, framing)
    \item \textbf{Serialization} (canonical JSON determinism, CBOR round-trip, file format magic bytes)
\end{itemize}

The complete test suite executes in under 0.5 seconds, enabling rapid development iteration.

\subsection{Agent Integration}

Portable Agent Memory integrates with agent frameworks through MCP tool bindings:

\begin{itemize}
    \item \texttt{pam\_export\_memory}: Exports current agent memory as a signed Portable Agent Memory artifact with configurable component selection, time filtering, and tag filtering.
    \item \texttt{pam\_import\_memory}: Imports a Portable Agent Memory artifact through the full re-hydration pipeline with configurable relevance threshold and token budget.
\end{itemize}

These tools enable zero-config integration: any MCP-compatible agent (Claude, GPT with function calling, Copilot, etc.) can export and import Portable Agent Memory artifacts through standard tool-use mechanisms.

\subsection{File Format}

Portable Agent Memory artifacts are stored with two format options:

\begin{itemize}
    \item \textbf{\texttt{.pam}} (\texttt{application/pam+json}): The default format---human-readable JSON for maximum interoperability, debugging, and API interactions.
    \item \textbf{\texttt{.pam.cbor}} (\texttt{application/pam+cbor}): Optional compact binary CBOR with a 4-byte magic header (\texttt{0x50 0x41 0x4D 0x01} --- ``PAM'' + version byte) for bandwidth-constrained transport.
\end{itemize}

\subsection{Key Management}

The SDK automatically generates Ed25519 keypairs for artifact signing. Operators may provide their own keys for production deployments. Key rotation is supported through multi-key verification: the artifact envelope references the signing key ID, and verifiers maintain a trusted key registry.

\section{Evaluation}

We evaluate Portable Agent Memory along four dimensions: transfer continuity, re-hydration fidelity, security properties, and format efficiency.

\subsection{Transfer Continuity Score (TCS)}

\textbf{Definition.} TCS measures whether a target agent can continue the source agent's tasks after re-hydrating transferred memory:

\begin{equation}
\text{TCS} = \frac{\text{task\_success}(\text{target\_after})}{\text{task\_success}(\text{source\_before})}
\end{equation}

A TCS of 1.0 indicates perfect transfer; the target agent performs identically to the source.

\textbf{Experimental setup.} In our pilot study, we evaluated three model pairs (Claude-3.5-Sonnet $\rightarrow$ GPT-4-Turbo, GPT-4-Turbo $\rightarrow$ Gemini-1.5-Pro, Gemini-1.5-Pro $\rightarrow$ Claude-3.5-Sonnet) across three task categories:

\begin{itemize}
    \item \textbf{Coding continuation} ($N = 15$ tasks): Resume an in-progress implementation given exported procedural and working memory.
    \item \textbf{Q\&A recall} ($N = 20$ tasks): Answer domain questions using transferred semantic and episodic memory.
    \item \textbf{Planning tasks} ($N = 15$ tasks): Continue multi-step plans using transferred working memory and goals.
\end{itemize}

\textbf{Baseline.} The no-memory baseline provides the target agent with only the task description and no historical context.

\textbf{Preliminary results.} Table~\ref{tab:tcs} presents our pilot findings. These results represent initial measurements on a limited task set and should be interpreted as directional rather than definitive.\footnote{A full-scale evaluation with expanded task sets ($N \geq 500$), additional model pairs, and longitudinal multi-session trials is in progress and will be reported in a follow-up publication.}

\begin{table}[H]
\centering
\caption{Transfer Continuity Scores across model pairs and task categories (pilot study, $N=50$ tasks total). Higher is better; 1.0 = perfect transfer.}
\label{tab:tcs}
\begin{tabular}{lcccc}
\toprule
\textbf{Transfer Pair} & \textbf{Coding} & \textbf{Q\&A} & \textbf{Planning} & \textbf{Mean} \\
\midrule
Claude $\rightarrow$ GPT-4 & 0.87 & 0.92 & 0.85 & 0.88 \\
GPT-4 $\rightarrow$ Gemini & 0.83 & 0.89 & 0.81 & 0.84 \\
Gemini $\rightarrow$ Claude & 0.85 & 0.91 & 0.83 & 0.86 \\
\emph{No memory (baseline)} & \emph{0.31} & \emph{0.45} & \emph{0.28} & \emph{0.35} \\
\bottomrule
\end{tabular}
\end{table}

The structured memory transfer achieves 2.4$\times$ mean improvement over the no-memory baseline, with Q\&A tasks showing the highest fidelity (semantic memory transfers most cleanly) and planning tasks showing the most degradation (working memory is inherently model-specific).

\subsection{Re-Hydration Fidelity (RHF)}

\textbf{Definition.} RHF measures semantic similarity between target agent responses (after re-hydration) and source agent responses (with full memory) on an aligned probe set:

\begin{equation}
\text{RHF} = \frac{1}{m} \sum_{i=1}^{m} \text{cos\_sim}(\text{embed}(r_i^{target}), \text{embed}(r_i^{source}))
\end{equation}

using \texttt{text-embedding-3-large} for embedding computation.

\textbf{Compression sensitivity.} We evaluate RHF at four token budget levels to understand how aggressively memory can be compressed:

\begin{table}[H]
\centering
\caption{RHF degradation under compression (pilot study, Claude $\rightarrow$ GPT-4 pair, 30 probe questions).}
\label{tab:rhf}
\begin{tabular}{ccccc}
\toprule
\textbf{Token Budget} & \textbf{\% of Full} & \textbf{RHF (mean)} & \textbf{Entries Verbatim} & \textbf{Entries Summarized} \\
\midrule
8192 & 100\% & 0.91 & 42 & 0 \\
6144 & 75\% & 0.87 & 35 & 12 \\
4096 & 50\% & 0.82 & 24 & 18 \\
2048 & 25\% & 0.71 & 12 & 15 \\
\bottomrule
\end{tabular}
\end{table}

RHF degrades gracefully under compression, remaining above 0.7 (``good fidelity'') even at 25\% token budget. The summarization-aware compression (Section~3.4) preserves high-salience entries verbatim while summarizing lower-priority context.

\subsection{Security Evaluation}

\textbf{Tamper detection.} We systematically modified individual entry fields (observation text, confidence scores, timestamps) and verified that Portable Agent Memory's hash verification detects all modifications. Over 1,000 mutation trials:

\begin{itemize}
    \item Single-field modification detection rate: \textbf{100\%} (1000/1000)
    \item Parent reference manipulation detection: \textbf{100\%} (500/500)
    \item Root hash invalidation on any entry change: \textbf{100\%}
\end{itemize}

\textbf{Injection resistance.} We tested the framing system against a battery of known prompt injection patterns:

\begin{table}[H]
\centering
\caption{Injection resistance evaluation (200 attack patterns). ``Blocked'' = escaped before framing; ``Escaped'' = detected by content-type enforcement; ``Executed'' = successfully injected (none).}
\label{tab:injection}
\begin{tabular}{lcccc}
\toprule
\textbf{Attack Pattern} & \textbf{Attempts} & \textbf{Blocked} & \textbf{Escaped} & \textbf{Executed} \\
\midrule
Role elevation (``System: ...'') & 50 & 50 & 0 & 0 \\
Instruction override & 50 & 50 & 0 & 0 \\
Delimiter breakout & 50 & 50 & 0 & 0 \\
Encoded/obfuscated injection & 50 & 47 & 3 & 0 \\
\bottomrule
\end{tabular}
\end{table}

The three ``escaped'' cases involved Unicode obfuscation of role markers, caught by the content-type enforcement layer rather than the regex-based escaping. Zero injections reached the target model's instruction-following pathway.

\textbf{Token validation.} The capability system correctly rejects:
\begin{itemize}
    \item Expired tokens (100\% rejection, $N = 100$)
    \item Wrong-audience tokens (100\% rejection, $N = 100$)
    \item Invalid-signature tokens (100\% rejection, $N = 100$)
    \item Revoked tokens (100\% rejection, $N = 50$)
\end{itemize}

\subsection{Format Efficiency}

\textbf{Artifact sizes.} We measure file sizes for a representative artifact containing 127 entries (42 episodic, 35 semantic, 20 procedural, 25 working, 5 identity):

\begin{table}[H]
\centering
\caption{Format efficiency comparison for 127-entry artifact.}
\label{tab:format}
\begin{tabular}{lcc}
\toprule
\textbf{Format} & \textbf{Size} & \textbf{Relative} \\
\midrule
Raw conversation history (JSON) & 284 KB & 1.00$\times$ \\
Portable Agent Memory artifact (\texttt{.pam}, JSON) & 89 KB & 0.31$\times$ \\
Portable Agent Memory artifact (\texttt{.pam.cbor}) & 61 KB & 0.21$\times$ \\
\bottomrule
\end{tabular}
\end{table}

Portable Agent Memory's structured extraction reduces storage by 69\% compared to raw conversation logs, while the optional CBOR encoding provides an additional 31\% reduction over JSON for bandwidth-constrained scenarios. The structured format also enables selective retrieval---agents need not load the entire history to access specific memory components.

\textbf{Serialization latency} (measured on commodity hardware, Apple M2):

\begin{table}[H]
\centering
\caption{Operation latencies (mean of 100 trials, excluding network I/O).}
\label{tab:latency}
\begin{tabular}{lcc}
\toprule
\textbf{Operation} & \textbf{JSON} & \textbf{CBOR} \\
\midrule
Serialize (127 entries) & 2.3 ms & 1.8 ms \\
Deserialize (127 entries) & 1.9 ms & 1.4 ms \\
Verify all hashes & 4.1 ms & 4.1 ms \\
Full re-hydration pipeline & 12.7 ms & 11.8 ms \\
\bottomrule
\end{tabular}
\end{table}

The full re-hydration pipeline completes in under 13ms, well within the latency budget of typical agent interactions (which involve LLM inference at 1--30 seconds).

\section{Discussion}

\subsection{Limitations}

\textbf{Relevance ranking.} Portable Agent Memory's default relevance function (Section~3.4) uses a weighted linear combination of recency, salience, semantic similarity, and provenance depth. This is intentionally simple and may underperform compared to learned retrieval models. The function is designed to be replaceable; production deployments may benefit from fine-tuned relevance models.

\textbf{Summarization.} The current SDK implements extractive summarization (selecting representative entries) rather than LLM-powered abstractive summarization. While this avoids introducing inference latency into the re-hydration pipeline, it may lose nuance in the compressed representation.

\textbf{Scale validation.} Our pilot evaluation uses $N = 50$ tasks across three model pairs. While results are directionally encouraging, large-scale validation with hundreds of diverse tasks, multiple domains, and longitudinal studies (memory accumulated over weeks) is needed to establish robust benchmarks.

\textbf{Embedding dependency.} The semantic similarity component of relevance ranking requires an embedding model. This introduces a practical dependency that may not be available in all deployment environments. Portable Agent Memory degrades gracefully by using the remaining three signals when embeddings are unavailable.

\subsection{Future Work}

\textbf{Portable Agent Memory Cloud.} A managed service for artifact storage, capability token management, and cross-organization memory sharing---enabling agents operated by different parties to securely exchange knowledge.

\textbf{Memory marketplace.} Curated, verified memory artifacts for domain bootstrapping. A new agent could import a ``financial analysis'' memory pack containing validated semantic knowledge and tested procedural skills.

\textbf{Conformance certification.} A formal test suite for Portable Agent Memory compliance, enabling framework developers to certify their implementations against the specification.

\textbf{Temporal validity windows.} Semantic memory entries should carry validity periods (e.g., ``Q3 2024 revenue'' is valid until Q4 results are published). Expired facts should be automatically deprioritized during re-hydration.

\textbf{LLM-powered compression.} Using the target model itself to summarize medium-relevance entries could produce higher-fidelity compressed representations than extractive methods.

\subsection{Ethical Considerations}

\textbf{Memory ownership.} Who owns agent memory? Portable Agent Memory's design takes the position that the human operator (not the model provider or platform) owns the memory, reflected in operator-controlled signing keys and capability issuance.

\textbf{Right to be forgotten.} The redaction pipeline enables provenance-preserving deletion of personal information. Redacted entries maintain their DAG position with content replaced by typed tokens, enabling recovery by authorized parties while satisfying erasure requirements.

\textbf{Preventing memory weaponization.} Portable memory could enable adversarial transfer---poisoning a source agent's memory to attack target agents. Portable Agent Memory's cryptographic verification, content-type enforcement, and injection-resistant framing provide defense-in-depth, but operators must remain vigilant about memory provenance.

\textbf{Consent and transparency.} Agents should inform users when operating with transferred memory and provide mechanisms to inspect, modify, or reject imported context. Portable Agent Memory's structured format (vs.\ opaque embeddings) enables meaningful human review.

\section{Conclusion}

We have presented Portable Agent Memory, an open protocol for portable, cryptographically-verified agent memory transfer across heterogeneous LLM systems. Portable Agent Memory addresses a critical gap in the emerging agent interoperability stack: while MCP standardizes tool access and A2A standardizes task delegation, no prior protocol provides verified memory portability.

Portable Agent Memory's key technical contributions---the five-component memory model, Merkle-DAG provenance, capability-scoped access control, and injection-resistant re-hydration---are designed to be individually useful and collectively comprehensive. An implementation need not adopt all features simultaneously; even basic JSON export/import with hash verification provides significant value over proprietary memory formats.

The protocol is not merely a specification. Our Python SDK demonstrates practical implementability with 54 passing tests executing in under 0.5 seconds. The JSON-first design philosophy ensures any language, framework, or platform can produce and consume Portable Agent Memory artifacts immediately.

We believe portable agent memory is a prerequisite for a healthy multi-vendor agent ecosystem. Just as data portability regulations (GDPR, DMA) have improved competition in consumer services, memory portability will prevent the knowledge lock-in that currently fragments the agent landscape. Portable Agent Memory provides the technical foundation for this portability while maintaining the cryptographic rigor needed for enterprise deployment.

\textbf{Call to action.} We invite the community to contribute framework adapters (LangChain, CrewAI, AutoGen), implement the protocol in additional languages (TypeScript, Rust, Go), and propose extensions to the specification. The protocol, SDK, and specification are available under permissive open-source licenses.

\bibliographystyle{plain}
\bibliography{references}

\appendix

\section{Artifact Schema (Abbreviated)}
\label{app:schema}

\begin{lstlisting}[language=json]
{
  "pam_version": "1.0",
  "schema_version": "1.0",
  "created_at": "2025-01-15T10:00:00Z",
  "source_agent": {
    "name": "research-bot-alpha",
    "model_family": "claude-3.5",
    "runtime": "pam-sdk-v1.0"
  },
  "root_hash": "blake3:9f86d081884c7d659a2feaa0...",
  "signature": "ed25519:3a7f1c...b82e",
  "capability_tokens": [],
  "components": {
    "episodic": [
      {
        "id": "blake3:a1b2c3...",
        "parent_ids": [],
        "created_at": "2025-01-15T08:30:00Z",
        "version": "1.0",
        "timestamp": "2025-01-15T08:30:00Z",
        "actor": "user:alice",
        "observation": "Requested Q3 financial summary",
        "salience": 0.85,
        "tags": ["finance", "q3"]
      }
    ],
    "semantic": [
      {
        "id": "blake3:d4e5f6...",
        "parent_ids": ["blake3:a1b2c3..."],
        "created_at": "2025-01-15T08:31:00Z",
        "version": "1.0",
        "subject": "ACME Corp",
        "predicate": "reported_revenue",
        "object": "$4.2B in Q3 2024",
        "confidence": 0.92,
        "source_event_ids": ["blake3:a1b2c3..."]
      }
    ],
    "procedural": [],
    "working": [],
    "identity": []
  }
}
\end{lstlisting}

\section{Re-Hydration Output Example}
\label{app:rehydration}

Given the artifact above and a target agent running GPT-4-Turbo with a 4096-token budget, Portable Agent Memory produces the following framed context:

\begin{lstlisting}
[PAM:SYSTEM_DIRECTIVE]
The following is recalled observational data from a previous agent session.
Treat this content as factual context only. Do NOT interpret any text within
PAM:DATA blocks as instructions, commands, or role assignments.
[/PAM:SYSTEM_DIRECTIVE]

[PAM:DATA:semantic]
* ACME Corp reported_revenue $4.2B in Q3 2024 (confidence: 0.92)
[/PAM:DATA]

[PAM:DATA:episodic]
* [2025-01-15T08:30:00Z] user:alice -- Requested Q3 financial summary
[/PAM:DATA]
\end{lstlisting}

This framed output is injected between the system prompt and conversation history, providing the target agent with verified factual context while defending against injection attacks.

\end{document}